# Irrelevance of $^1$H composition to the superconductivity in the infinite-layer nickelates: judging from the MeV energy scale

Jia-Cai Nie[1,2]*, Xing-Yu Chen[1], Yi Bian[3], Xue-Yan Wang[1], Ting-Na Shao[1], Jing-Xin Gao[3], Wei Mao[4]*, Bing-Hui Ge[5], Arnold Müller[6], and Jikun Chen[3]*

[1] School of Physics and Astronomy, Beijing Normal University, Beijing 100875, China
[2] Key Laboratory of Multiscale Spin Physics, Ministry of Education, Beijing Normal University, Beijing 100875, China
[3] School of Materials Science and Engineering, University of Science and Technology Beijing, Beijing 100083, China
[4] School of Engineering, The University of Tokyo, 2-11-16 Yayoi, Bunkyo-ku, Tokyo 113-0032, Japan
[5] Institutes of Physical Science and Information Technology, Anhui University, Hefei, Anhui, 230601, China
[6] Laboratory of Ion Beam Physics, ETH Zurich, CH-8093 Zurich, Switzerland

Correspondence: Prof. Jia-Cai Nie (jcnie@bnu.edu.cn), Dr. Wei Mao (maowei@iis.u-tokyo.ac.jp), and Prof. Jikun Chen (jikunchen@ustb.edu.cn).

**Abstract**

The discovery of the superconductivity in the infinite-layer nickelates, as topotactically reduced from their respective perovskite percussors via co-annealing with $CaH_2$, extends the understanding in superconductivity. Nevertheless, whether the incorporated $^1$H composition is critical to the infinite-layer superconductivity recently arouses considerable debates, while the central challenge lies in the quantification of $^1$H that is easily interfered by the conventional electron or orbital associated processes. Herein, we demonstrate the irrelevance between the superconductivity in the infinite-layer nickelates and their incorporated $^1$H composition, assisted by nuclear reaction analysis (NRA) and heavy ion energy recoil detection analysis (HIERDA) based on the nuclear interactions at MeV energy scale. These approaches completely overwhelm the conventional interferes, such as ionization, activation and chemical bonds, and achieves the $^1$H quantification within superconducting $La_{0.8}Sr_{0.2}NiO_2$ (or $Nd_{0.8}Sr_{0.2}NiO_2$). A large diversity of $^1$H composition far beyond the previously expected critical dome was observed, while their $T_C$ were not changed significantly. Furthermore, the superconductivity was demonstrated to be achievable for $La_{0.8}Sr_{0.2}NiO_2$ reduced by Al without any hydrogen associated process, while the superconducting properties for the $CaH_2$ reduced $La_{0.8}Sr_{0.2}NiO_2$ is rather stable after long term exposure in air, despite the high volatility of $^1$H within oxides. All these results indicate that the $^1$H incorporation composition is not critical to the superconductivity of the infinite-layer nickelates.

## 1. Introduction

Since the discovery of high-temperature superconductivity in cuprates[1], scientists have been striving to elucidate the physical mechanism of high-temperature superconductivity and search for new superconductors with higher $T_C$ (even room temperature levels) or new mechanisms (e.g., p-wave or interfacial superconductivity)[2-7]. The discovery of infinite-layer nickelate superconductivity[8] has sparked a new round of research interest. The parent material of nickelate superconductivity has a similar crystal structure to one of the parent materials of cuprate high-temperature superconductivity, and its outermost orbital electron configuration ($Ni^{1+}$ $3d^9$) is similar to that of parent materials of many cuprate superconductors[9-11]. Therefore, in-depth exploration and research on relevant nickelates can help us further understand the physical origins of unconventional high-temperature superconductivity. To achieve the $Ni^{1+}$ configuration, a perovskite precursor (*e.g.*, $Nd_{0.8}Sr_{0.2}NiO_3$ or $La_{0.8}Sr_{0.2}NiO_3$) was firstly deposited, followed by a soft chemical reduction process by annealing together with $CaH_2$[8]. During such process, the $^1H$ composition is inevitably incorporated within as formed infinite-layer nickelates. The hydrogen interactions with correlated transitional metal (*TM*) oxides were previously demonstrated to passivate defects[12-14], donate carriers[15,16], or even directly reshaping orbital configuration and electron occupancy[17-20]. In particular for the perovskite nickelates, huge regulations in the material resistivity were previously achieved via the hydrogenation triggers Mottronic transitions among the complicated electronic phase diagrams, such as the electron more itinerant $Ni^{3+}$ $t^6_{2g}e^1_g$, electron localized $Ni^{2+}t^6_{2g}e^2_g$[17], and even the superconducting $t^6_{2g}e^3_g$[8,21]. Nevertheless, the role of the $^1H$ composition to the superconductivity in infinite-layer nickelates is yet under debate[22-25].

Recently, Ding *et al* demonstrated that the hydrogen composition was critical for the superconductivity of the infinite-layer nickelates[22]. Nevertheless, such claim was soon doubted by Balakrishnan *et al*[23], Gonzalez *et al*[24] and Di Cataldo *et al*[25], from the perspective of theoretical calculations and analogous experimental investigations. Ding *et al* tempted to quantify $^1H$ by probing the relative amount of the secondary $^1H^-$ from the nickelate samples compared to the one from the $KAl_2[AlSi_3O_{10}](OH)_2$ reference using time-of-flight-secondary-ion mass spectrometry (ToF-SIMS), while soon afterwards Balakrishnan *et al*[23] demonstrated that the major $^1H$-containing secondary ion under the same SIMS condition should be $^{16}O^1H^-$ rather than $^1H^-$. Also, Gonzalez *et al*[24] recently report that no significant increase in hydrogen density between the perovskite and infinite-layer phases was observed by SIMS, and this indicates the irrelevance between the $^1H$ incorporation and the infinite-layer nickelate superconductivity. It is worth noticing that SIMS was generally recognized as a semi-quantitative approach as both the composition and kinetic energy ($E_K$) distribution of the secondary ions are rather sensitive to the chemical atmosphere and surface conductivity of the samples, and this challenges the reliability and central claim in Ding's work.

Herein, we achieved hydrogen quantification for infinite-layer nickelate superconducting samples, by using nucleus associated interactions (*e.g.*, nuclear reaction or elastic collision) at MeV energy scale, such as nuclear reaction analysis (NRA)[26] and elastic recoil detection analysis with heavy ions (HI-ERDA)[27]. In MeV energy scale, the electron-involved processes below keV (*e.g.*, excitation, ionization, chemical bonding and dissociation) that interfere conventional $^1$H quantification were completely overwhelmed. The $^1$H composition within superconducting $H_xLa_{0.8}Sr_{0.2}NiO_2$ and $H_xNd_{0.8}Sr_{0.2}NiO_2$, as reduced by $CaH_2$ soft chemical treatments, was found largely beyond the expected critical superconducting dome, compared to Ding's report[22]. Furthermore, similar superconductivity was also achieved in $La_{0.8}Sr_{0.2}NiO_2$ through aluminothermic reduction without introducing any hydrogenation process, while the superconducting properties of $La_{0.8}Sr_{0.2}NiO_2$ reduced by $CaH_2$ remain unchanged after 8 months. These results indicate the irrelevance between the superconductivity associated with the infinite-layer nickelates and their incorporated $^1$H composition.

2. Results and Discussions

**Nuclear interactions within MeV energy scale enables $^1$H quantification:** Quantifying the $^1$H (or $^1$H$^+$) absolute compositions within solids based on conventional electron (or orbits) associated principles below MeV is challenging. The $^1$H exhibits only a single K-shell electron and interacts weakly with the common probing electrons or X-ray, and this impedes its direct detection via conventional approaches, such as X-ray diffraction (XRD), X-ray Photoelectron Spectroscopy (XPS), and Auger electron spectroscopy (AES).[26] Although the $^1$H (or $^1$H$^+$) can be sputtered out of the solids and probed via secondary ion mass spectrometry (SIMS) as the working principle illustrated in Figure 1a, it is more likely to achieve a qualitative or semiquantitative $^1$H detection. This is owing to the complexity in the composition of the hydrogen containing secondary ions, the absolute detection amount of which is known to be easily interfered by the conductivity and/or chemical atmospheres of the samples. As has been already pointed by Balakrishnan et al[23], complex charges (e.g., H$^+$ and H$^-$) and chemical fragments (e.g., at least OH$^-$) were observed for the $^1$H-containing secondary ions during the SIMS measurement of the infinite layer nickelates. Furthermore, it is also worth noticing that the respective secondary neutral species (e.g., H and OH) must be coexist together with the secondary ions as sputtered out by the analysis beam. In Figure 1a, we also list the complicated interconversion relationship among the multiple $^1$H containing species as sputtered out by the analysis beam, including (at least) the ionization or de-ionization, electron attaching or de-attaching, dissociation and combination, and excitation (de-excitation). It is worth noticing that their respective interconversion energies are at the same range compared to the ones associated to the chemical bonds within solids (e.g., $10^{-1}$-$10^1$ eV). Therefore, vary either the sample or the incident beam are expected to change the interconversions among these H-containing secondary species, and this impedes quantifying the absolute $^1$H composition by just detecting one specific $^1$H-containing secondary ion using SIMS.

To achieve a strict $^1$H quantification, it is vital to elevate the interaction energy to a high enough magnitude that completely overwhelms the conventional interferes associated to the electronic structure of the sample materials (e.g., charge, orbital and chemical bonds). Therefore, the nucleus-associated high energy (e.g., $10^0$-$10^1$ MeV) processes, such as the nuclear reaction (e.g., in NRA) and elastic recoil collisions (e.g., in ERDA), were previously applied to achieve accurate $^1$H composition. As illustrated in Figure 1b, the NRA can directly probe the $^1$H within solid, regardless of its valance and chemical atmosphere, based on the nuclear reaction between $^1$H and $^{15}$N at a sharp resonant kinetic energy ($E_K$) of ~6.385 MeV that release the characteristic

$\gamma$-ray[26]. As a constant energy loss is expected when the $^{15}N^{2+}$ incident ion penetrates the material, the $^1H$ depth profile can be further obtained by increasing the incident $E_K$ of $^{15}N^{2+}$ stepwise from ~6.385 MeV. Alternatively, $^1H$ quantification can also be achieved using ERDA based on its elastic recoil collisions by $^{127}I^+$ incident ions at MeV range (e.g., herein 13 MeV)[27], as illustrated in Figure 1c. In MeV range the collision between the incident beam and the atom in the material can be viewed as elastic, and the atomic mass within the material is indicated by the ratio between the kinetic energy of the recoiled species compared to the incident energy at a fixed detection geometry. The depth profile can be also obtained as the energy loss of the recoiled species is constant that results in the reduction in the recoiled energy with an increasing distribution depth. Thus, by further considering the collision cross-section of the different elements within the material, the stoichiometry can be obtained.

**Growth and electronic structure of the infinite-layer nickelates:** For judging the relevance of $^1H$ composition to the superconductivity of nickelates using the above MeV associated techniques, we grew the $La_{0.8}Sr_{0.2}NiO_3$ and $Nd_{0.8}Sr_{0.2}NiO_3$ precursor films on $SrTiO_3(001)$ substrates using pulsed laser deposition (PLD) at an oxygen pressure of 19 Pa and substrate temperature of 600 °C, according to our previous reports[28,29]. Five $La_{0.8}Sr_{0.2}NiO_3$ films (S-1 to S-5) and one $Nd_{0.8}Sr_{0.2}NiO_3$ film (S-7) were reduced to infinite-layers by using $CaH_2$ soft chemical treatments at 320 °C (S-1 to S-5 and S-7), while one $La_{0.8}Sr_{0.2}NiO_3$ was reduced *in-situ*[6] by depositing an Al layer at 200 °C on its surface (S-6). The variations in their crystal structures before and after the above reductions by either $CaH_2$ or Al are demonstrated by their XRD patterns as compared in Figure 2a for $La_{0.8}Sr_{0.2}NiO_3$ ($La_{0.8}Sr_{0.2}NiO_2$), while more XRD patterns for the infinite-layers are further shown in Figure S1. It clearly demonstrates the same variation in crystal structure from the perovskite to infinite-layer when reducing $La_{0.8}Sr_{0.2}NiO_3$ by either $CaH_2$ or Al. Figure 2b further demonstrates the representative cross-section morphology for the infinite-layers, and it can be seen that the $La_{0.8}Sr_{0.2}NiO_2$ film coherently grown on the $SrTiO_3$ substrate at a thickness of ~9 nm.

To probe the variation in electronic structures owing to the reduction processes, the synchrotron-based X-ray absorption spectrum (XAS) analysis was performed for the $La_{0.8}Sr_{0.2}NiO_3$ precursor, the *ex-situ* $CaH_2$ reduced $La_{0.8}Sr_{0.2}NiO_2$, and the *in-situ* Al-reduced $La_{0.8}Sr_{0.2}NiO_2$. As demonstrated in Figure 2c, the relative intensity of rightwards shoulder peak of Ni-$L_3$ edge was reduced after either $CaH_2$ or Al reduction, while the Ni-$L_2$ edge shift leftwards. Meanwhile, the reduction reduces the intensity of the pre-peak in O-$K$ edge, as shown in Figure 2d. These observations indicate the similar tendency in the respective variations in the electronic structure owing to the reduction in the valance state of Ni via either $CaH_2$ or Al reduction. As further shown in Figure S2, the relative intensity of the peak B compared to peak A in the Ni-$L_3$ edge is observed to be slightly lower for $La_{0.8}Sr_{0.2}NiO_2$ as *ex-situ* reduced by $CaH_2$ than the one *in-situ* reduced by Al, indicating a lower valance of Ni of the $CaH_2$ reduced sample compared to the Al reduced one.

**Comparing the superconductivity for initial judgement:** To characterize their superconductive properties, we measured the temperature dependence of resistivity ($\rho$-$T$ tendency) for the seven samples, as shown in Figure 3a. It can be seen that all samples, no matter whether reduced by $CaH_2$ or Al, exhibit superconducting transition similar to the previous reports[29-33]. We further calculated their critical temperatures (e.g., $T_{C,50\%R}$, $T_{C,10\%R}$, and $T_{C,90\%R}$), and the method is similar to the previous reports[22], more details are shown in Figure S3 and Table S1. As the Al reduction does not involve any hydrogen (or hydrogenation) associated process, it challenges the previous expectation that the hydrogen composition was critical to the superconductivity of the infinite-layer nickelates.

The relevance between the infinite-layer nickelate superconductivity and its expected critical $^1H$ composition was further judged from the stability in their superconducting properties. Noticing that the $^1H$ (or proton) are highly volatile, the properties of oxides that are indeed related to the $^1H$ composition are expected to degrade after long term exposure to air. This was previously observed for the hydrogenated vanadium dioxide ($H_xVO_2$), in which situation the resistivity of the $^1H$ associated electron-localized $H_xVO_2$ fast reduced in air, e.g., by more than three orders within one day[34]. Analogously, if the superconductivity of $H_xNd_{0.8}Sr_{0.2}NiO_2$ was only achievable within a narrow dome (e.g., x=0.225-0.275) according to Ding's work[22], the superconductivity of the $CaH_2$ reduced $La_{0.8}Sr_{0.2}NiO_2$ would likely to be jeopardized after long term exposure to air. In Figure 3b, the $\rho$-$T$ tendencies of the $CaH_2$ reduced $La_{0.8}Sr_{0.2}NiO_2$ were compared before and after exposure in air for 8 months. Surprisingly, the superconductivity of the $CaH_2$ reduced sample is rather stable and was not degraded after long term exposure in air, and this observation also disproves the previous claim about the critical role in hydrogen composition to the superconductivity of nickelates.

**$^1H$ compositions probed by NRA and HI-ERDA:** To achieve strict $^1H$ composition ($x$) for the superconducting samples, both approaches of NRA (in the University of Tokyo) and HI-ERDA (in ETH, Zurich) were used. Figure 4a shows the depth profile of $^1H$ within $La_{0.8}Sr_{0.2}NiO_2$ or $Nd_{0.8}Sr_{0.2}NiO_2$ as measured by NRA, while the ones measured by HI-ERDA are demonstrated in Figure S4. The NRA spectra is composed of two parts: 1) a surface peak associated with both the surface resonance and surface absorbed $^1H$), and 2) a bulk distribution as proportion to the $^1H$ within the film (or substrate). Considering the thickness of the film, the $^1H$ was averaged from the penetration depth of 5-8 nm, where the surface resonance peak already dissipate without reaching the substrate. It can be seen that a large diversity in the $^1H$ composition, e.g., some $CaH_2$ reduced samples exhibit rather low $^1H$ composition approaching to the Al-reduced one or the background signal, while much higher $^1H$ compositions (e.g., beyond $5\times10^{21}$ $cm^{-3}$) are also possible. As the $^1H$ is rather volatile within oxides, many extrinsic factors (e.g., the sample transports from Beijing to Tokyo or Beijing to Zurich) are expected to result in diversity in their $^1H$ composition. Also, it is interesting to note that the $^1H$ is also possible to diffuse into the substrate just underneath the film, in particular, for the ones with high $^1H$ composition. Similar phenomenon was also observed by Balakrishnan *et al*[23] in SIMS measurements, as high amount of $OH^-$ signal was detected in the substrate region for the heavily hydrogenated nickelate films in SIMS measurements.

**Irrelevant $^1H$ compositions to superconductivity:** In Figure 4b, the $T_C$ ($T_{C,90\%R}$, $T_{C,50\%R}$ and $T_{C,10\%R}$) is plotted as a function of the hydrogen composition, comparing to the expected superconducting dome as report in Ding's work[22]. We have to clarify that as-presented tendency at least holds when the $^1$H composition was measured, as superconducting properties of these samples were not changed much before and after the NRA measurements, despite some radiation damage that elevate the resistivity a bit. It can be seen that the tendencies of both the $T_{C,90\%R}$-$x$ and $T_{C,10\%R}$-$x$ herein largely fall beyond the expected superconducting dome as reported by Ding *et al*[22], while there is even no dome-like tendency observed for $T_C$ with their $^1$H compositions. This further confirms the irrelevance between the infinite-layer superconductivity to certain critical $^1$H compositions, in contradiction to Ding's recent claim.

It is worth mentioning that the diversity in the $^1$H compositions and depth distributions as observed for the superconducting infinite-layer nickelates (also the case in previous reports[22,23]) are likely to be caused by the high diffusivity and volatility of $^1$H (or $^1$H$^+$) within oxides. From the one aspect, during the hydrogen annealing, the $^1$H incorporation concentration can be easily elevated by extrinsically introducing interfacial or lattice defects, as $^1$H is known to passivate the respective dangling bonds and likely to be aggregated around the defect enriched regions[12-14,21]. From the other aspect, the incorporated $^1$H (or $^1$H$^+$) composition within the hydrogenated samples is expected to be gradually 'escaped' out of the film, even these samples are well preserved, e.g., in vacuum. In particular, such $^1$H loss within $La_{0.8}Sr_{0.2}NiO_2$ (or $Nd_{0.8}Sr_{0.2}NiO_2$) coherently grown on $SrTiO_3$ (001) is expected to be accelerated by their tensile distortion, similar to the situation as reported previously for the tensile strained perovskite nickelates (e.g., $SmNiO_3$)[35].

## 3. Conclusion

In conclusion, we demonstrate the irrelevance between the superconductivity as achieved for the infinite-layer nickelates and the $^1$H incorporation composition during their topotactic reduction process as co-annealed with $CaH_2$, from the following three aspects. Firstly, the superconductivity was demonstrated to be achievable for $La_{0.8}Sr_{0.2}NiO_2$ reduced by Al without any hydrogen associated process. Secondly, the superconducting property for the $CaH_2$ reduced $La_{0.8}Sr_{0.2}NiO_2$ is rather stable after long term exposure in air, despite the high volatility of $^1$H within oxides. Thirdly, a large diversity in $^1$H composition largely beyond the previously expected critical dome was observed in superconducting $La_{0.8}Sr_{0.2}NiO_2$ and $Nd_{0.8}Sr_{0.2}NiO_2$ using NRA and HIERDA, while their $T_C$ were not changed significantly. This clarification is important to prohibit the misleading of the direction from both perspectives of superconducting mechanism understanding and its potential applications.

**Competing interests:** We declare no competing financial interest.

**Additional information:** Supplementary Information is available for this manuscript.

**Correspondences:** Correspondence should be addressed: Prof. Jia-Cai Nie (jcnie@bnu.edu.cn), Dr. Wei Mao (maowei@iis.u-tokyo.ac.jp), and Prof. Jikun Chen (jikunchen@ustb.edu.cn).

**Acknowledgments:** This work was supported the National Key Research and Development Program of China (No. 2021YFA0718900) and the National Natural Science Foundation of China (92065110).

**Figures and caption**

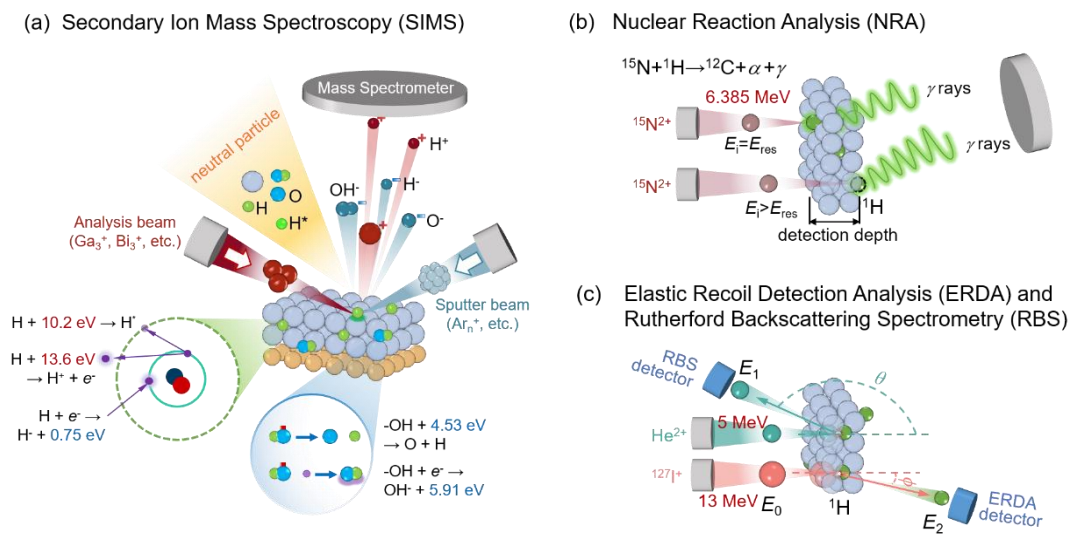

**Figure 1**. Comparing the working principle of **(a)** second ion mass spectrometry **(b)** nuclear reaction analysis (NRA) and **(c)** elastic recoil detection analysis (ERDA) for potential $^1$H detection. In SIMS measurements, the secondary species as sputtered out by the analysis beam exhibit complex compositions (e.g., not only ions but also neutral species), while the interconversion relationship among the $^1$H containing species occurs at a similar low energy range compared to the chemical bonds within solid (e.g., $10^{-1}$-$10^1$ eV). Therefore, the variation in chemical atmosphere and/or surface conductivity of the sample is expected to vary the absolute detection amount of a specific $^1$H containing ion that impedes a strict quantification of the $^1$H composition of the sample. In contrast, the NRA and ERDA utilize the nuclear associated interactions at MeV energy scale, in which case the electron associated interferes (e.g., chemical bonds or potential excitations) are completely overwhelmed. The NRA detects the characteristic gamma-rays as emitted from the nuclear reaction between the $^{15}$N$^{2+}$ incident ion from the accelerator and the $^1$H with regardless of its valance charge within the material that is triggered at a characteristic incident kinetic energy ($E_K$) of ~6.385 MeV. As a constant energy loss is expected when the $^{15}$N$^{2+}$ incident ion penetrates the material, the $^1$H depth profile is obtained by increasing the incident $E_K$ of $^{15}$N$^{2+}$ stepwise from ~6.385 MeV. In ERDA, the heavy incident ion (e.g., $^{127}$I$^+$) is used to cause the elastic recoil of the atoms within the material, while their recoil energy ($E_2$) compared to the initial incident energy (e.g., $E_0$=13 MeV) at a fixed detection geometry is probed that indicates the atomic mass of the incident ion and the sample.

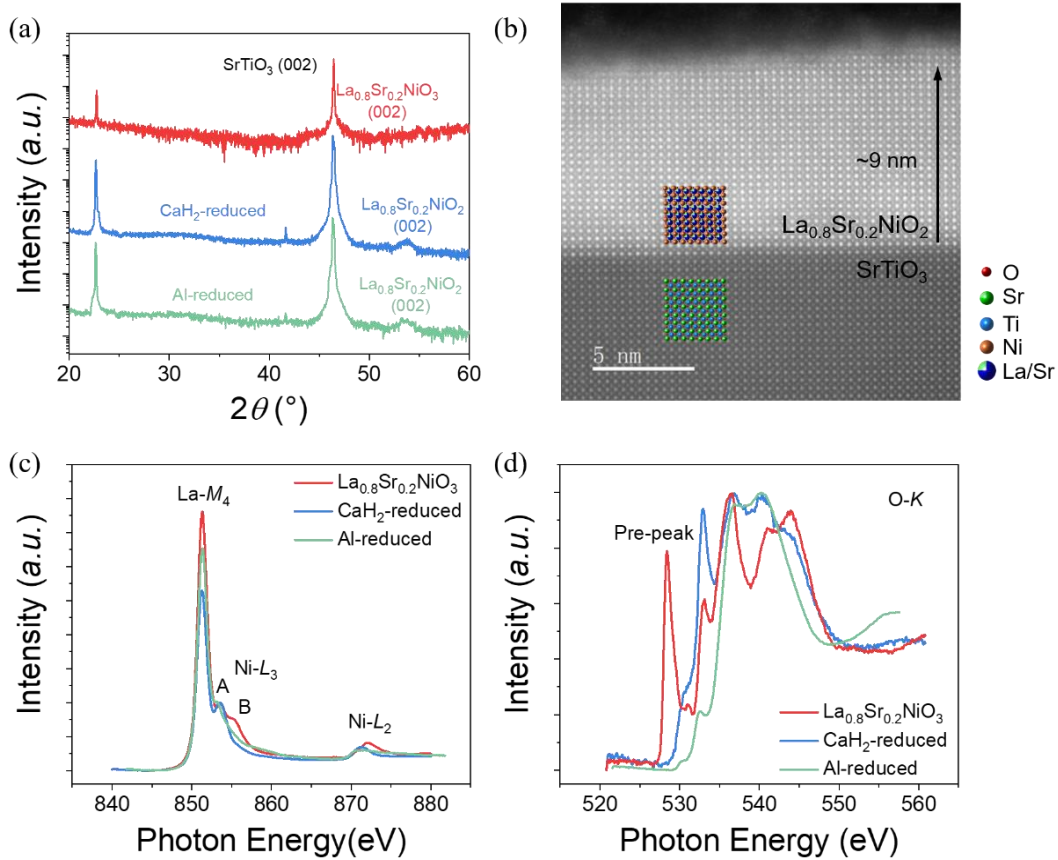

**Figure 2**. **(a)** Representative X-ray diffraction (XRD) patterns for as deposited perovskite structured La$_{0.8}$Sr$_{0.2}$NiO$_3$ and the infinite-layer La$_{0.8}$Sr$_{0.2}$NiO$_2$ as further reduced via *ex-situ* CaH$_2$ and *in-situ* Al. **(b)** Representative cross section morphology of the La$_{0.8}$Sr$_{0.2}$NiO$_2$/SrTiO$_3$ as reduced by CaH$_2$. **(c),(d)** The synchrotron-based XAS of the **(c)** Ni-*L* and **(d)** O-*K* edges for the La$_{0.8}$Sr$_{0.2}$NiO$_3$ precursor, the *ex-situ* CaH$_2$ reduced La$_{0.8}$Sr$_{0.2}$NiO$_2$, and the *in-situ* Al-reduced La$_{0.8}$Sr$_{0.2}$NiO$_2$.

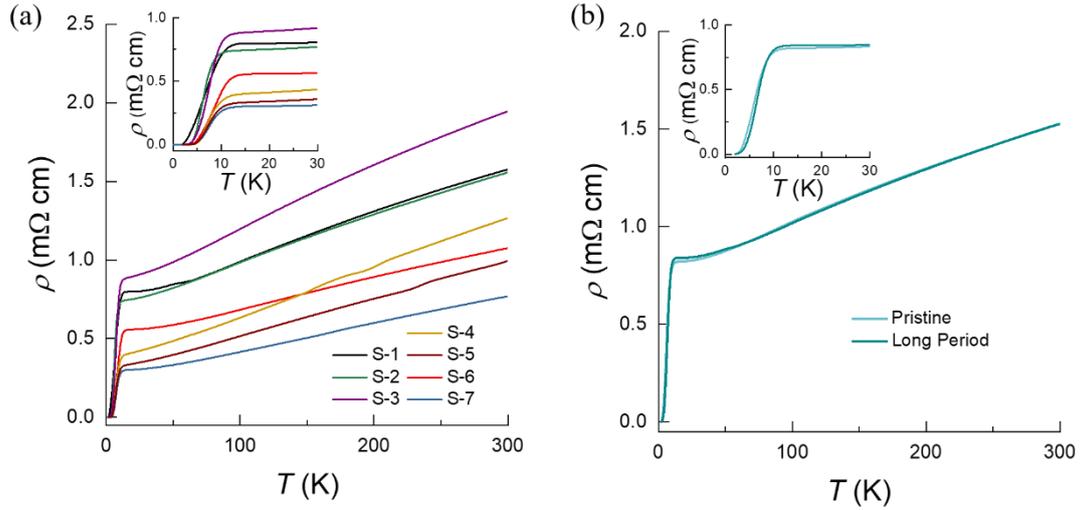

**Figure 3**. **(a)** Temperature dependence of resistivity ($\rho$-$T$) as plots for the $La_{0.8}Sr_{0.2}NiO_2$ films reduced by $CaH_2$ (S-1 to S-5), the $La_{0.8}Sr_{0.2}NiO_2$ film reduced by Al (S-6), and the $Nd_{0.8}Sr_{0.2}NiO_2$ film reduced by $CaH_2$ (S-7[29]). **(b)** The $\rho$-$T$ curves for the $La_{0.8}Sr_{0.2}NiO_2$ as measured immediately after $CaH_2$ reduction and after exposure in air for 8 months that demonstrates the stability of its superconductivity in air. The zoom in $\rho$-$T$ curves in the low temperature range are shown in the insets of **(a)** and **(b)**.

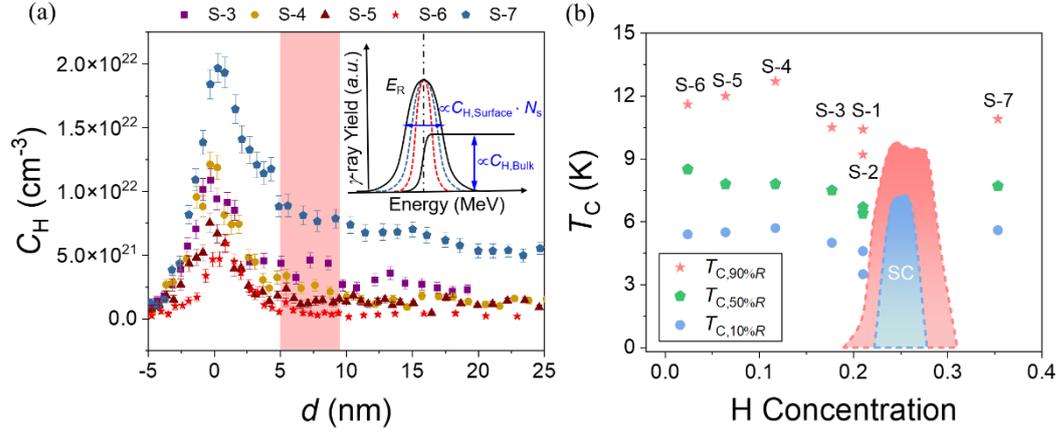

**Figure 4**. (a) The nuclear reaction analysis (NRA) spectra for the *ex-situ* $CaH_2$ reduced $La_{0.8}Sr_{0.2}NiO_2$ (S-3, S-4 and S-5) and $Nd_{0.8}Sr_{0.2}NiO_2$ (S-7), as well as the *in-situ* Al reduced $La_{0.8}Sr_{0.2}NiO_2$ (S-6). The inset illustrates the two parts in the NRA spectra: 1) a surface resonance peak, and (2) a bulk distribution. The intensity of the surface resonance peak at the resonance energy ($E_R$) is contributed by the surface resonance with the incident ion (not the presence of hydrogen) and is mainly Doppler broadened, while the presence of surface absorption of $^1H$ will result in further broadening of the surface resonance peak proportionally. Therefore, herein the $^1H$ concentration distributed within the thin film region of ~5-8 nm away from the surface (as marked in each figure) was averaged to get the $^1H$ composition, as marked by the color region. (b) Plotting the superconducting transition temperature ($T_C$) versus as-measured hydrogen composition ($x$) for the samples by HI-ERDA (S-1, S-2) and NRA (S3 to S-7) as shown in (a), compared to the dome-like tendency reported by Ding *et al*[22]. The red, green and blue symbols represent for $T_{C,10\%R}$, $T_{C,50\%R}$ and $T_{C,90\%R}$, respectively, as calculated the same to Ding's work, compared to the expected superconducting boundaries of $T_{C,10\%R}$ and $T_{C,90\%R}$, respectively, by Ding *et al*[22], as marked by the red and blue shadows.

**Supporting Information**

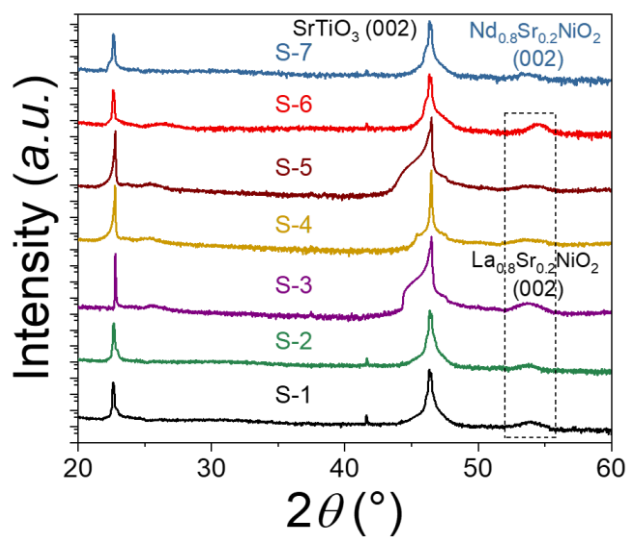

**Figure S1**. The X-ray diffraction (XRD) patterns for the *ex-situ* CaH$_2$ reduced La$_{0.8}$Sr$_{0.2}$NiO$_2$ (Sample-1 to Sample-5) and Nd$_{0.8}$Sr$_{0.2}$NiO$_2$ (Sample-7) and *in-situ* Al reduction (Sample-6). The diffraction peaks at a 2$\theta$ of ~53.9° associated with the (002) peaks of the infinite-layer are observed for all samples.

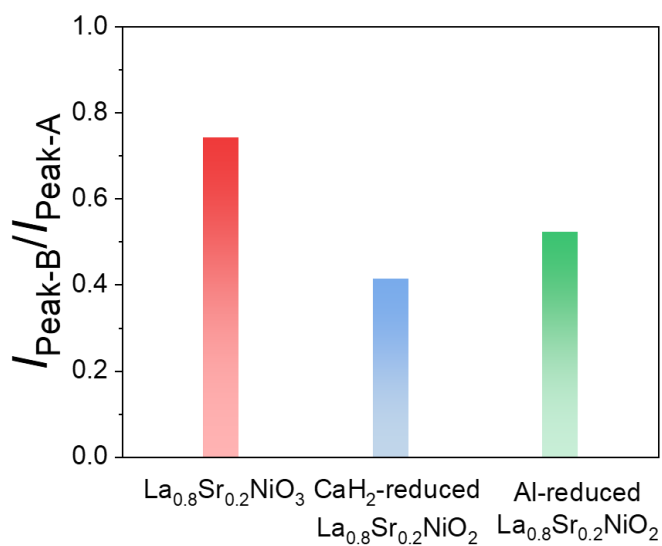

**Figure S2**. Comparing the ratio of the peak B to peak A in the Ni-$L_3$ edge of the X-ray absorption spectra shown in Figure 2c as measured for the La$_{0.8}$Sr$_{0.2}$NiO$_3$ precursor, the *ex-situ* CaH$_2$ reduced La$_{0.8}$Sr$_{0.2}$NiO$_2$, and the *in-situ* Al-reduced La$_{0.8}$Sr$_{0.2}$NiO$_2$.

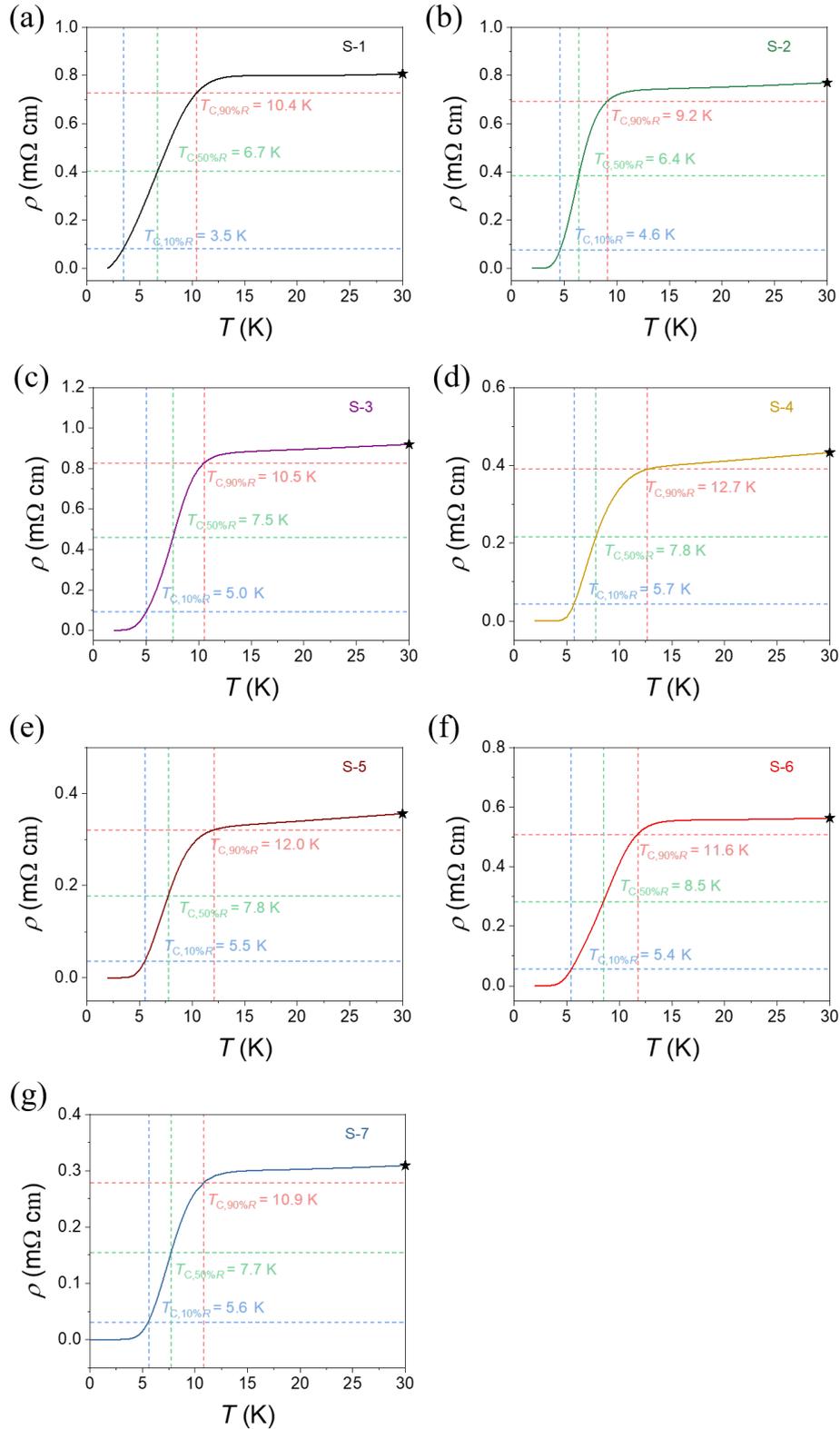

**Figure S3**. Determination of the $T_{C,10\%R}$, $T_{C,50\%R}$ and $T_{C,90\%R}$, associated with the superconducting properties of the La$_{0.8}$Sr$_{0.2}$NiO$_2$ or Nd$_{0.8}$Sr$_{0.2}$NiO$_2$ samples (S1-S7) as shown in Fig.4b. The $T_{C,10\%R}$, $T_{C,50\%R}$ and $T_{C,90\%R}$ are the critical temperatures when the relative resistivity decreases by 10%, 50% and 90%, respectively, compared with the resistivity at 30 K.

**Table S1.** Comparing the $T_{C,10\%R}$, $T_{C,50\%R}$ and $T_{C,90\%R}$, associated with the superconducting properties of the $La_{0.8}Sr_{0.2}NiO_2$ or $Nd_{0.8}Sr_{0.2}NiO_2$ samples (S1-S7) as shown in Fig.4b.

| Sample | $T_{C,10\%R}$ (K) | $T_{C,50\%R}$ (K) | $T_{C,90\%R}$ (K) |
|---|---|---|---|
| S-1 | 3.5 | 6.7 | 10.4 |
| S-2 | 4.6 | 6.4 | 9.2 |
| S-3 | 5.0 | 7.5 | 10.5 |
| S-4 | 5.7 | 7.8 | 12.7 |
| S-5 | 5.5 | 7.8 | 12.0 |
| S-6 | 5.4 | 8.5 | 11.6 |
| S-7 | 5.6 | 7.7 | 10.9 |

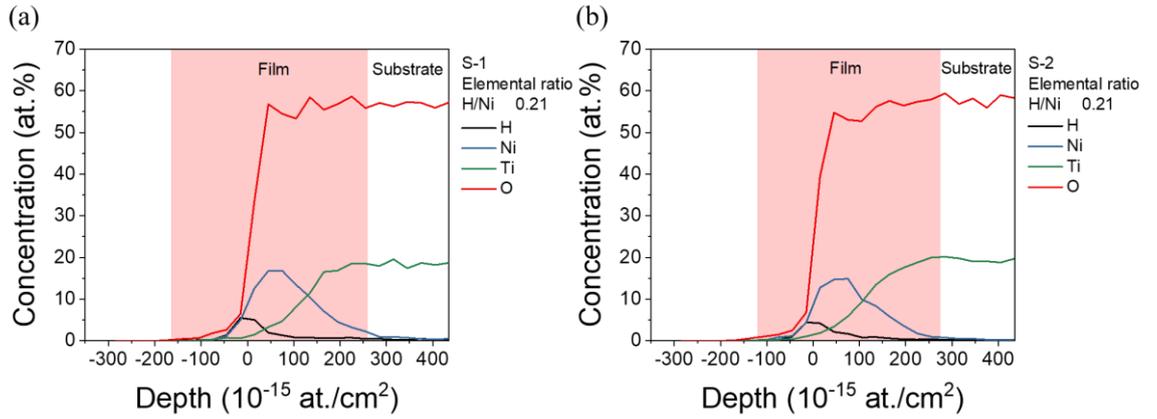

**Figure S4.** Depth profile of the relative concentration of $^1H$, Ni, Ti and O with the sample as measured for **(a)** S-1 and **(b)** S-2 using HI-ERDA.